\documentclass[prd,aps,preprintnumbers,amsmath,amssymb,nofootinbib]{revtex4}
\usepackage{bm} 
\usepackage{epsfig}

\begin{document}
\title{Constituent quarks, chiral symmetry, and chiral point of the constituent quark model} 
\author{Wolfgang Lucha$^{a}$, Dmitri Melikhov$^{a,b}$, and Silvano Simula$^{c}$} 
\affiliation{
$^a$ Institute for High Energy Physics, Austrian Academy of Sciences, Nikolsdorfergasse 18, A-1050, Vienna, Austria\\
$^b$ Institute of Nuclear Physics, Moscow State University, 119992, Moscow, Russia\\
$^c$ INFN, Sezione di Roma 3, Via della Vasca Navale 84, I-00146, Roma, Italy}
\date{\today}
\begin{abstract}
We construct the full axial current of the constituent quarks by a summation of the 
infinite number of diagrams describing constituent-quark soft interactions. 
By requiring that the conservation of this current is violated only by terms of order $O(M_\pi^2)$, 
where $M_\pi$ is the mass of the lowest pseudoscalar $\bar QQ$ bound state, 
we derive important constraints on (i) the axial coupling $g_A$ of the 
constituent quark and (ii) the $\bar QQ$ potential at large distances. 
We define the chiral point of the constituent quark model as those values of the parameters, 
such as the masses of the constituent quarks and the couplings in the $\bar QQ$ potential, 
for which $M_\pi$ vanishes. 
At the chiral point the main signatures of the spontaneously broken chiral symmetry are 
shown to be present, namely: the axial current of the constituent quarks is conserved, the leptonic 
decay constants of the excited pseudoscalar bound states vanish, and the pion decay constant 
has a nonzero value. 
\end{abstract}
\maketitle
\section{Introduction}
Chiral symmetry is a basic symmetry of massless QCD which leads to the conservation of 
the axial-vector current, apart from the axial anomaly in the singlet-flavor channel. 
The masses of the light $u$ and $d$ quarks are small compared 
to the confinement scale, and therefore the chiral limit serves as a good approximation 
for the light-quark sector of QCD. The consequences of chiral symmetry for the QCD Green 
functions in the non-perturbative region and the expansion of these functions in 
powers of the external momenta and the small quark masses have been worked out within chiral 
perturbation theory \cite{chpt}. Chiral symmetry in QCD is spontaneously broken, and 
therefore it is not a symmetry of the hadron spectrum: except for the existence of the 
octet of light pseudoscalars, the lowest part of the hadron spectrum shows no trace of 
chiral symmetry. 

Because of confinement, the calculation of the hadron mass spectrum directly from the QCD Lagrangian 
is a very challenging task, which requires a non-perturbative approach. 
For the description of the mass spectrum 
of hadrons and their interactions at low momentum transfers, QCD-inspired constituent quark models 
(i.e., models based on constituent-quark degrees of freedom in which mesons appear as $\bar QQ$ 
bound states in a potential) proved to be quite successful \cite{gi,gromes}. 
Moreover, there are many pieces of evidence that the constituent-quark picture provides a 
good description not only of the mass spectrum of hadrons, 
but also of their interactions at not too large momentum transfers 
\cite{anisbook,amn,ms,silvano}. Just because of the proper description of the hadron mass spectrum, 
the Lagrangian of the constituent quark model {\it cannot} be chirally invariant 
(otherwise it would produce a chirally invariant spectrum of hadron states).\footnote{
Several versions of chirally invariant Lagrangians based on massive constituent quarks have been 
discussed in the literature \cite{mg,se}. In this case the axial current constructed as the Noether 
current from the Lagrangian is explicitly conserved. However, explicit chiral symmetry 
for massive constituent quarks at the Lagrangian level requires the inclusion of Goldstones along with 
the constituent-quark degrees of freedom. Although very elegant, these approaches are not well suited for 
the description of the meson spectrum: in constituent-quark potential models, such as the Godfrey-Isgur 
model \cite{gi}, mesons are nicely described as bound states of constituent quarks, leaving no 
room for additional Goldstone degrees of freedom. Inclusion of both constituent quarks and
Goldstones leaves doubts on possible double counting of meson states in these approaches.}
As a result, the Noether axial current constructed in such models is not conserved. 

In this paper we show that, nevertheless, a properly formulated constituent quark model 
has a ``chiral point'' which corresponds to the chiral limit of QCD. 
We start by constructing the full axial current of the constituent quarks by a summation of the 
infinite number of diagrams describing constituent-quark soft interactions. 
By requiring that the conservation of this current is violated only by terms $O(M_\pi^2)$, 
we derive important constraints on the axial coupling of the constituent quark, $g_A$, and 
on the  $\bar QQ$ potential at large distances. 
We then define the chiral point as those values of the quark-model parameters 
(masses of the constituent quarks and couplings in the quark potential) for which the mass 
of the pseudoscalar $\bar QQ$ ground state vanishes. 
The following signatures of the spontaneously broken chiral symmetry may be observed at the chiral point: 
\begin{itemize}
\item[a.] The axial current of the constituent quarks is conserved. This requires a relation between 
the axial coupling of the constituent quark, $g_A(s)$, and the pion wave function $\Psi_\pi(s)$ 
in the chiral limit, which we derive explicitly. 
\item[b.] The decay constant of the pion remains finite. 
\item[c.] The decay constants of the excited massive pseudoscalars vanish. 
\item[d.] The coupling $g_A(s)$ should be finite at $s=0$ in accordance with a spontaneously 
broken chiral symmetry. This requires that the potential of the $\bar QQ$ interaction 
saturates at large distances: $V(r)\to {\rm const}$ as $r\to \infty$. 
\end{itemize}

The paper is organized as follows: 
In Section \ref{2}, we recall the basic properties of the axial current in QCD. In Section \ref{3}, 
we construct the full non-perturbative axial current of the constituent quarks by taking into 
account their soft interactions. We obtain the constraint on the axial coupling of the 
constituent quark, $g_A(s)$, which provides the conservation of the full axial current of 
the constituent quarks when the mass of the lowest pseudoscalar $Q\bar Q$ 
bound state vanishes. The constraints on the $Q\bar Q$ potential which lead to a  
non-vanishing $Q\bar Q \pi$ coupling in the chiral limit are derived. 
In Section \ref{4}, we discuss the properties of pseudoscalar mesons at the chiral point. 
Section \ref{5} gives our conclusions. 
Appendix \ref{app} gives the connection between the behavior of the $\bar QQ$ potential at large $r$ 
and the analytic properties of the bound-state wave function.  
In Appendix \ref{aapp} we discuss vector and scalar couplings of the constituent quarks.

\section{\label{2}Axial current in QCD}
Let us briefly recall the main properties of the axial current in QCD: 
the axial current $j^5_\alpha(x)=\bar u(x) \gamma_\alpha \gamma_5 d(x)$ and the pseudoscalar 
current $j^5(x)=i\bar u(x)\gamma_5 d(x)$ are related by   
\begin{eqnarray}
\label{div}
\partial^\alpha j^5_\alpha(x)=(m_u+m_d)j^5(x). 
\end{eqnarray}
The axial current is conserved in the limit of massless quarks (the
chiral limit). This leads to specific properties 
of its correlators \cite{lucha}. The coupling of a 
pseudoscalar meson to these currents has the form 
\begin{eqnarray}
\label{fpf5}
\langle 0|\bar u\gamma_\alpha\gamma_5 d|P(q)\rangle&=&if_P q_\alpha, \qquad 
\langle 0|i\bar u\gamma_5 d|P(q)\rangle={f^5_P}. 
\end{eqnarray}
The divergence equation (\ref{div}) requires 
\begin{eqnarray}
\label{relation}
f_PM_P^2\propto m,  
\end{eqnarray}
implying that at least one of the quantities on the l.h.s.\ vanishes in the chiral limit. 
If chiral symmetry is spontaneously broken, Eq.~(\ref{relation}) leads to the following 
alternatives \cite{svz}: 
\begin{eqnarray}
\label{constraints}
&&M^2_\pi=O(m),\quad f_\pi=O(1),\quad \mbox{ground-state pion} \nonumber\\  
&&M^2_{P}=O(1),\quad f_{P}=O(m),\quad \mbox{excited pseudoscalars}.  
\end{eqnarray}
Note that the non-vanishing of the pion decay constant in the chiral limit means that 
the generator of the axial symmetry $Q^5=\int d\vec x\, \bar q\gamma_0\gamma_5 q(x)$ does not 
annihilate the vacuum but rather produces a massless pion from the vacuum state.  
Thus, the vacuum is not invariant under chiral transformations, and chiral symmetry is spontaneously broken. 
If no spontaneous breaking of chiral symmetry occurs, then in the chiral limit the pion 
behaves the same way as the excited pseudoscalars: it stays massive and its decay constant 
vanishes \cite{krassnigg}.
The divergence equation (\ref{div}) leads to the following relation between the couplings:  
\begin{eqnarray}
\label{relP}
M_P^2 f_P  ={(m_u+m_d)}f^5_P. 
\end{eqnarray}
For the pion, by virtue of the Gell-Mann--Oakes--Renner (GMOR) formula \cite{gmor} 
\begin{eqnarray}
\label{gmor}
f_\pi^2 M^2_\pi =-(m_u+m_d)\langle \bar uu+ \bar dd\rangle+O(M_\pi^4), 
\end{eqnarray}
we obtain 
\begin{eqnarray}
\label{relpi}
f_\pi^5 =-\frac{\langle \bar uu+ \bar dd\rangle}{f_\pi}. 
\end{eqnarray}

\section{\label{3}Conserved non-perturbative axial current of the constituent quarks}
In this section we construct the full axial current of the constituent quarks 
by summing soft interactions among the latter. 
We show that the current obtained 
by this procedure is conserved if the spectrum of the pseudoscalar $\bar QQ$ bound states 
contains a massless state. 

\subsection{The constituent-quark interaction amplitude}
Let us start with a discussion of the amplitude of the constituent $Q\bar Q$ interaction. 
We are interested in the region of small invariant mass of the $Q\bar Q$ pair, and 
we want to take into account only two-particle intermediate $Q\bar Q$ states. 
In the region of small invariant mass of the $\bar QQ$ pair, 
the $J^P=0^-$ partial S-wave $Q\bar Q$ amplitude, $A$, which satisfies the two-particle unitarity condition, 
may be parameterised in the form \cite{anisovich}
\begin{eqnarray}
\label{1.1}
A=
\frac{\bar Q i\gamma_5 Q}{\sqrt{N_c}}
\frac{\bar Q i\gamma_5 Q}{\sqrt{N_c}}
\,\frac{G^2(p^2)}{1-B(p^2)}, 
\end{eqnarray}
with the function $G(p^2)$ having no singularities in the region $p^2>0$.
The two-particle unitarity relation leads to the expression 
\begin{eqnarray}
\label{1.2}
B(p^2)=\frac{1}{\pi}\int\frac{ds}{s-p^2}G^2(s)\rho(s), 
\end{eqnarray}
with 
\begin{eqnarray}
\label{1.3}
\label{rhop}
\rho(s)&=&\frac{1}{N_c}{\rm Im} 
\left[i \int d^4xe^{ipx}\langle 0|T(\bar Q i\gamma_5 Q(x),\bar Q i\gamma_5 Q(0))|0\rangle \right]
\nonumber\\ 
&=&\frac{s-(m_1-m_2)^2}{8\pi s}\lambda^{1/2}(s,m_1^2,m_2^2)\theta(s-(m_1+m_2)^2), 
\end{eqnarray}
where $\lambda(s,m_1^2,m_2^2)=(s-m_1^2-m_2^2)^2-4 m_1^2m_2^2$ is the triangle function. 
By a formal expansion of the denominator in (\ref{1.1}), 
the amplitude $A$ may be represented by the series of diagrams 
\begin{eqnarray}
\label{fig:1}
\epsfig{file=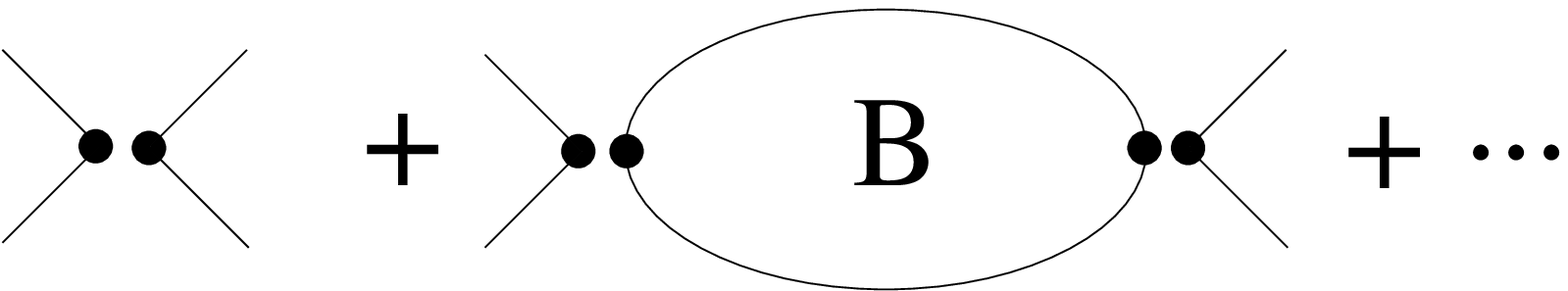,height=.8cm}
\end{eqnarray}
with small solid circles denoting $G(p^2)$. 
These diagrams may be generated by the nonlocal vertex \cite{anisovich,melikhov}
\begin{eqnarray}
\label{1.4}
\label{bareamp}
\frac{\bar Q(-k_1) i\gamma_5 Q(k_2)}{\sqrt{N_c}}\cdot 
\frac{\bar Q(-k'_1) i\gamma_5 Q(k'_2)}{\sqrt{N_c}}\, {G^2(p^2)}, 
\qquad p=k_1+k_2=k'_1+k'_2.  
\end{eqnarray}
The pseudoscalar meson corresponds to a pole in the amplitude $A$, and its mass $M_P$ is obtained from 
\begin{eqnarray}
\label{1.5}
1-B(M_P^2)=0.  
\end{eqnarray}
The parameterization (\ref{1.1}) corresponds to a separable Ansatz for the $N$-function 
of the $N/D$-representation
of the $\bar QQ$ partial-wave scattering amplitude, which allows one to describe the 
interaction of the $\bar QQ$ bound state with external currents 
(see \cite{anisovich} and references therein). 
This simple Ansatz clearly has a limited applicability: namely, it is suitable only for 
the low-energy region and leads to the appearance of only one --- lowest-mass --- 
$\bar QQ$ bound state. 
Nevertheless, the amplitudes of the interaction of this bound state with the external currents, obtained with 
the nonlocal separable vertex (\ref{1.4}), satisfy rigorous requirements of gauge 
invariance and analyticity. 
This approximation is very convenient for constructing the full axial current of constituent quarks, 
and can be easily generalized to include excited states \cite{anisovich}. 

\subsection{Axial current of the constituent quarks}
The interaction of the constituent quarks with gluons is constructed through their 
covariant derivatives, 
and the constituent-quark currents satisfy the following divergence equation: 
\begin{eqnarray}
\label{divQ}
\partial^\mu (\bar Q(x) \gamma_\mu \gamma_5 Q(x))&=&2m_Q \bar Q(x) i \gamma_5 Q(x). 
\end{eqnarray}
Therefore, the partially conserved axial current of the constituent quarks, similar to the case of the
axial current of the nucleon, should contain not only the $\bar Q \gamma_\mu\gamma_5 Q$ structure 
but also the induced pseudoscalar term $\bar Q \gamma_5 Q$: 
\begin{eqnarray}
\label{1.6a}
\langle 0|\bar q \gamma_\mu \gamma_5 q |\bar QQ\rangle&=&
\bar Q \left\{g_A(p^2)\gamma_\mu \gamma_5 + g_P(p^2)\, p_\mu\, \gamma_5\right\} Q. 
\end{eqnarray}
In the chiral limit the conservation of the current (\ref{1.6a}) leads to the relation 
$g_P(p^2)=\frac{2m_Q}{p^2}g_A(p^2)$ \cite{berthold}. The form factor $g_P(0)$ contains a pole at $p^2=0$ if 
$g_A(0)\ne 0$. 

We shall see that the pseudoscalar term in the full axial current of the constituent quark emerges after 
taking into account the soft interactions of the constituent quarks described by the
amplitude (\ref{1.1}). Moreover, we shall see that this term contains a pole at $p^2=M_\pi^2$ and 
that the axial current (\ref{1.6a}) is conserved in the chiral limit. 

We start with the axial-vector structure of the constituent-quark current 
(the Noether current obtained from the Lagrangian of the constituent quark model), 
which we refer to as the bare current:  
\begin{eqnarray}
\label{1.6}
\langle 0|\bar q \gamma_\mu \gamma_5 q |\bar QQ\rangle_{\rm bare}&=&
g_A(p^2) \bar Q \gamma_\mu \gamma_5 Q. 
\end{eqnarray}
As known from the application of the constituent quark model to light mesons and baryons \cite{ss}, 
the coupling $g_A(p^2)$ is a slowly varying function of $p^2$, and the 
``on-shell'' axial coupling of the 
constituent quark is close to unity, $g_A(4m_Q^2)\simeq 1$ \cite{weinberg}. 

Let us take into account the soft interactions generated by the vertex (\ref{1.4})\footnote{Quark 
interactions in the $J^P=1^-$ channel also contribute to the full axial current. These interactions may be
generated by the vertex 
$\bar Q \Gamma_\alpha\gamma_5 Q\cdot  \bar Q \Gamma_\alpha\gamma_5 Q\, G^2_A(p^2)$, where the operator 
$\bar Q \Gamma_\alpha \gamma_5 Q$ is constructed to be orthogonal to $\bar Q \gamma_5 Q$, 
i.e., ${\rm Sp} (\bar Q \Gamma_\alpha\gamma_5 Q \bar Q \gamma_5 Q)=0$. 
Inclusion of this structure leads to the appearance of an additional {\it transverse} term 
in the full expression for the axial current of the constituent quarks. 
This term contains poles corresponding to axial mesons, and is irrelevant for the region of small 
$p^2$ we are interested in. Therefore, we shall not take this spinorial structure into account in our analysis.}. 
Retaining only two-particle $\bar QQ$ singularities, 
as done already for the amplitude (\ref{1.1}), 
the full axial current is given by the set of diagrams 
\vspace{.4cm}
\begin{eqnarray}
\label{fig:2}
\epsfig{file=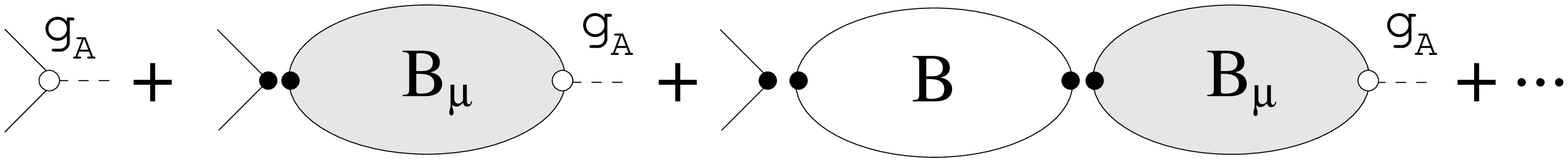,height=0.8cm}
\end{eqnarray}
where small solid circles denote $G$ and small empty circles denote $g_A$. 
The first term in this series corresponds to the bare current given by 
Eq.~(\ref{1.6}). 
The loop diagram $B_\mu$ has the following expression \cite{melikhov}: 
\begin{eqnarray}
\label{1.7}
B_\mu=ip_\mu B_{A}(p^2), \qquad B_{A}(p^2)=\frac{1}{\pi}\int\frac{ds}{s-p^2}G(s)g_A(s)\rho_{A}(s), 
\end{eqnarray}
with $\rho_{A}$ defined by 
\begin{eqnarray}
\label{1.8}
p_\mu \rho_{A}(p^2)=\frac1{\sqrt{N_c}}{\rm Im} \left[i\int d^4x\,e^{ipx}
\langle 0| T(\bar Q i\gamma_5 Q(x),\bar Q \gamma_\mu \gamma_5 Q(0))|0\rangle\right].  
\end{eqnarray}
Explicit calculations give
\begin{eqnarray}
\label{1.9}
\rho_{A}(s)=2m_Q\sqrt{N_c}\frac{\rho(s)}{s}. 
\end{eqnarray}
Making use of this relation, we obtain
\begin{eqnarray}
\label{1.9aa}
B_{A}(p^2)=2m_Q \sqrt{N_c}\int\frac{ds}{\pi\,s(s-p^2)}G(s)g_{A}(s){\rho(s)}. 
\end{eqnarray}
This expression may be written in the form 
\begin{eqnarray}
\label{1.10}
B_{A}(p^2)=2m_Q\sqrt{N_c}\frac{1}{p^2}\left\{\tilde B(p^2)-\tilde B(0)\right\}, 
\end{eqnarray}
with  
\begin{eqnarray}
\label{1.9ab}
\tilde B(p^2)=\int\frac{ds}{\pi(s-p^2)}G(s)g_A(s) \rho(s). 
\end{eqnarray}
Notice that $\tilde B$ reduces to $B$ if one replaces $g_A(s)$ by $G(s)$. 

Summation of the diagrams in (\ref{fig:2}) leads 
to the following expression for the axial current of the constituent quarks:
\begin{eqnarray}
\label{1.11}
\langle 0|\bar q \gamma_\mu \gamma_5 q |\bar QQ\rangle&=&
g_A(p^2) \bar Q \gamma_\mu \gamma_5 Q
+
\frac{\bar Q i\gamma_5 Q}{\sqrt{N_c}}\;G(p^2) \frac{i p_\mu B_{A}(p^2)}{1-B_{}(p^2)}.
\end{eqnarray}
Making use of Eq.~(\ref{1.10}), we obtain 
\begin{eqnarray}
\label{axial}
\label{1.12}
\langle 0|\bar q \gamma_\mu \gamma_5 q |\bar QQ\rangle&=&
g_A(p^2)  
\bar Q \gamma_\mu \gamma_5 Q
+
2m_Q \frac{p_\mu}{p^2}
\bar Q \gamma_5 Q \; G(p^2)\frac{\tilde B_{}(0)-\tilde B_{}(p^2)}{1-B_{}(p^2)}, 
\end{eqnarray}
which may be recast in the following convenient form:  
\begin{eqnarray}
\label{axial1}
\langle 0|\bar q \gamma_\mu \gamma_5 q |\bar QQ\rangle&=&
g_A(p^2) \left\{\bar Q \gamma_\mu \gamma_5 Q
+2m_Q \frac{p_\mu}{p^2}\bar Q \gamma_5 Q\right\}\nonumber\\
&&
+2m_Q \frac{p_\mu}{p^2}\bar Q \gamma_5 Q
\frac{g_A(p^2)\left(B(p^2)-1\right)
+G(p^2)\left(\tilde B(0)-\tilde B(p^2)\right)}{1-B(p^2)}.
\end{eqnarray}
The term in curly brackets is conserved by virtue of Eq.~(\ref{divQ}). 
We now require the second term to be of order 
$O(M_\pi^2)=O(m)$ in accordance with the divergence of the axial current in QCD. 
To provide for such a behavior, the constituent-quark axial coupling $g_A$ cannot be a 
constant, but should depend on the momentum as follows: 
\begin{eqnarray}
\label{ga1}
g_A(s)=\eta_A \,G(s)+O(M_\pi^2),  
\end{eqnarray}
with constant $\eta_A$. 
This relation leads to the following relation between the functions $B$ and $\tilde B$: 
\begin{eqnarray}
\label{ss}
\tilde B(p^2)=\eta_A \,B(p^2)+O(M_\pi^2).   
\end{eqnarray}
The pion corresponds to the pole in the amplitude (\ref{axial1}), which implies
\begin{eqnarray}
\label{1.14}
B(p^2=M_\pi^2)=1.  
\end{eqnarray}
Expanding $B(p^2)$ and $B(0)$ near $B(M_\pi^2)$ in Eq.~(\ref{axial1}), and making use of 
Eq.~(\ref{ss}), the axial current takes the form 
\begin{eqnarray}
\label{1.15}
\langle 0|\bar q \gamma_\mu \gamma_5 q |\bar QQ\rangle&=&
g_A(p^2) \left\{\bar Q \gamma_\mu \gamma_5 Q
+2m_Q \frac{p_\mu}{p^2}\bar Q \gamma_5 Q\right\}+
2m_Q g_A(p^2) \frac{p_\mu}{p^2}\bar Q \gamma_5 Q\,\frac{O(M_\pi^2)}{p^2-M_\pi^2}. 
\end{eqnarray}
Thus, taking into account soft interactions among constituent quarks leads to a 
conserved axial current if the mass spectrum of the model contains a massless pseudoscalar. 

It should be recalled that the spontaneous breaking of chiral symmetry requires that for a 
massive fermion (such as a nucleon or a constituent quark) 
the coupling $g_A(s)$ does not vanish for
$s=M_\pi^2$. 
As we have found, the conservation of the axial current at the chiral point requires that 
$g_A(s)=\eta_A G_\pi(s)+O(M_\pi^2)$. Therefore, to be compatible with the 
spontaneous breaking of chiral symmetry, the potential model should lead to the 
light pseudoscalar bound state for which $G_\pi(s=M_\pi^2)=\mbox{const}\ne 0$. 
The vertex $G_\pi(s)$ is related to the radial wave function by \cite{anisovich,melikhov}
\begin{eqnarray}
\Psi_\pi(s)=\frac{G_\pi(s)}{s-M_\pi^2}. 
\end{eqnarray} 
Therefore the condition $G_\pi(M_\pi^2)\ne 0$ implies that $\Psi_\pi$ should have a pole at $s=M_\pi^2$.

As shown in Appendix \ref{app}, in order a pole in $\Psi_\pi(s)$ to occur at $s=M_\pi^2$ , 
the potential of the $Q\bar Q$ interaction should saturate at large $r$:
\begin{eqnarray}
\label{saturation}
V(r\to\infty)={\rm const}. 
\end{eqnarray}
In this case the nearly massless pion is a strongly bound $Q\bar Q$ state with the 
binding energy $\epsilon\simeq 2m$. 
\newpage

\subsection{Pseudoscalar current of the constituent quarks}
Similarly, starting with the bare pseudoscalar current
\begin{eqnarray}
\langle 0| \bar q\gamma_5 q|\bar QQ\rangle_{\rm bare}&=&g_5(p^2)\bar Q \gamma_5 Q,
\end{eqnarray}
the effect of soft quark interactions leads to the full pseudoscalar current 
\begin{eqnarray}
\label{j_P}
\langle 0| \bar q\gamma_5 q|\bar QQ\rangle&=&g_5(p^2) \bar Q \gamma_5 Q\frac{1}{1-B(p^2)}.
\end{eqnarray}
Making use of the QCD divergence equation (\ref{div}) and the divergence equation for the constituent
quarks (\ref{divQ}), we obtain from Eqs.~(\ref{j_P}) and (\ref{axial})
\begin{eqnarray}
(m_u+m_d)g_5(p^2)=2m_Q g_A(p^2)\left[B(M_\pi^2)-B(0)\right]+O(M_\pi^2)=
2m_Q g_A(p^2)M_\pi^2 B'(0)+O(M_\pi^2).
\end{eqnarray}
The terms denoted as $O(M_\pi^2)$ emerge from the $O(M_\pi^2)$ terms in Eq.~(\ref{ga1}). 
If they are numerically small, then in the chiral limit $g_5(p^2)\propto g_A(p^2)$, and  
by virtue of the GMOR relation we find 
\begin{eqnarray}
\label{r5}
\frac{g_5(p^2)}{g_A(p^2)}=-2m_Q B'(0)\frac{\langle \bar uu+ \bar dd \rangle}{f_\pi^2}. 
\end{eqnarray}
The vector and scalar couplings of the constituent quarks are considered in Appendix
\ref{aapp}.

\section{\label{4}Chiral point of the constituent quark model}
We now consider certain properties of pseudoscalar mesons 
making use of the results obtained in the previous section. 

\subsection{Decay constants of pseudoscalar mesons}
We start with the decay constants $f_P$ and $f^5_P$ defined in (\ref{fpf5}) for both 
the ground-state and the excited pseudoscalar mesons. 
Isolating the pole term at $p^2=M_P^2$ in the representations of the 
axial and the pseudoscalar current gives the following expressions 
for the axial and pseudoscalar couplings of a pseudoscalar meson \cite{melikhov}:\footnote{
In the non-relativistic limit, $g_A(s)\to 1$: then Eq.~(\ref{fp}) 
is reduced to the standard non-relativistic relation $f_P(n)=\sqrt{{12}/{M_P(n)}}\Psi_n(\vec r=0)$.}  
\begin{eqnarray}
\label{fp}
f_{P}(n)&=&
\sqrt{N_c}\int 
ds g_A(s)\Psi_n(s)\rho(s,m_Q^2,m_Q^2)\frac{2m_Q}{s},
\\
\label{f5p}
f^5_{P}(n)&=&
\sqrt{N_c}\int ds g_5(s) \Psi_n(s)\rho(s,m_Q^2,m_Q^2), 
\end{eqnarray}
with the wave functions of the pseudoscalar mesons, $\Psi_{n}$, normalized according to 
\begin{eqnarray}
\label{orthonorm}
\int 
ds \Psi_{n}(s)\Psi_{m}(s)\rho(s,m_Q^2,m_Q^2)=\delta_{mn}. 
\end{eqnarray}
In the region $s\ge 4m^2$, the function $g_A(s)$ may be expanded over the full system of the 
eigenfunctions 
$\Psi_n(s)$. To provide for the conservation of the axial current in the chiral limit $M_\pi=0$, the expansion should have the 
following functional form: 
\begin{eqnarray}
\label{ga}
g_A(s)=\eta_A \Psi_0(s)(s-M_\pi^2)+M_\pi^2\sum_{n=0}^\infty  C_n \Psi_n(s), \qquad  C_n=O(1).
\end{eqnarray}
Substituting this expression into (\ref{fp}) we find
\begin{eqnarray}
f_{P}(n)&=&2m_Q \eta_A \sqrt{N_c}
\int ds \Psi_{0}(s)\Psi_{n}(s)\rho(s,m_Q^2,m_Q^2)\frac{s-M_\pi^2}{s}+O(M_\pi^2).  
\end{eqnarray}
For the ground state, $n=0$, the decay constant is clearly finite in the chiral limit. 
Making use of Eq.~(\ref{orthonorm}) gives the relation 
\begin{eqnarray}
f_\pi=2m_Q\eta_A \sqrt{N_c}+O(M_\pi^2). 
\end{eqnarray}
For excited states, $n\ne 0$, by virtue of the orthogonality condition (\ref{orthonorm}), we find 
\begin{eqnarray}
f_{P}(n\ne 0)&=&-2m_Q \eta_A M_\pi^2
\int ds \Psi_{0}(s)\Psi_{n}(s)\frac{\rho(s,m_Q^2,m_Q^2)}{s}+O(M_\pi^2). 
\end{eqnarray}
This decay constant is proportional to $M_\pi^2$ and thus vanishes in the chiral limit. Also beyond the chiral limit,  
the decay constants of the excited pseudoscalars are predicted to be much suppressed compared to the 
pionic decay constant.  
However, we cannot give further predictions for the decay constants of the excited pseudoscalars, 
since in this case the unknown terms $\sim M_\pi^2$ are of the same order as the contribution given 
by the main term in $g_A(s)$. A better knowledge of the details of $g_A(s)$ is necessary.\footnote{
The terms $O(M_\pi^2)$ in $g_A(s)$, Eq. (\ref{ga1}), cannot be obtained within the constituent quark picture itself,
but may be determined, e.g., by comparing the results for the excited pseudoscalars obtained from constituent quark
picture with those from other approaches (e.g. lattice QCD, Schwinger-Dyson equation, etc.). Presently, only a few
results for the first excited state are available, which are unfortunately not sufficient for a reliable
determination of these $O(M_\pi^2)$ terms.}

For the pseudoscalar coupling $f^5_{P}(n)$, making use of the expression (\ref{r5}), we obtain 
\begin{eqnarray}
f^5_{P}(n)&=&\eta_5
\sqrt{N_c}\int ds \Psi_{0}(s)\Psi_{n}(s)\rho(s,m_Q^2,m_Q^2)(s-M_\pi^2)+O(M_\pi^2), \qquad 
\eta_5 =-\eta_A 2m_Q B'(0)\frac{\langle \bar uu+ \bar dd \rangle}{f_\pi^2}.  
\end{eqnarray}
This coupling does not vanish in the chiral limit both for the ground and the excited states, 
in accordance with (\ref{relP}). 

Let us notice that the usual approximation, $g_A(s)={\rm const}$, may work well,  
at least for those quantities which do not vanish in the chiral limit, for the following
reason: In the relativistic 
quark model \cite{anisovich,melikhov}, the observables are given by integrals over $s$ 
along the two-particle cut. 
The region of $s$ near the two-particle threshold $s=4m^2$ is suppressed by the two-particle 
phase space, which vanishes at the threshold. The region of large $s$ is suppressed by the wave 
function. So the main contribution comes from intermediate values of $s$, 
where the specific details of the function $g_A(s)$ are not essential. 
The approximation of a constant $g_A(s)$ may even work numerically for the decay constants 
of the excited pseudoscalars for the physical values of the quark masses.  
However, this approximation cannot be applied for studying the behavior of these 
quantities in the chiral limit. 

\subsection{Pionic coupling of hadrons}
The expression for the full axial current (\ref{axial}) contains an explicit pion pole, 
thus providing the possibility to extract the amplitude of the pionic decay  
$h_1\to h_2 \pi$: 
\begin{eqnarray}
\label{amp}
p_\mu A(h_1\to h_2\pi)
=\lim_{p^2\to M_\pi^2}\frac{p^2-M_\pi^2}{f_\pi}\langle h_2|j^5_\mu|h_1\rangle
=p_\mu \frac{2m_Q}{f_\pi}\langle h_2|\bar Q\gamma_5 Q|h_1\rangle. 
\end{eqnarray}
It is understood that the amplitude $\langle h_2|\bar Q\gamma_5 Q|h_1\rangle$ is calculated 
making use of the constituent quark description of the hadrons $h_1$ and $h_2$. 
The expression (\ref{amp}) for the amplitude has been successfully applied to pionic decays of 
charmed mesons \cite{charm}. 

\subsection{The chiral constituent quark mass}
We give now an estimate for the constituent quark mass corresponding to the chiral limit, 
$m_Q^0$, making use of the following relation between the constituent quark mass 
$m_Q$ and the current quark mass $m$ at the chiral-symmetry breaking scale $\mu_\chi\simeq 1$ GeV \cite{ms2004}: 
\begin{eqnarray}
\label{ms}
\langle \bar qq\rangle =\frac{N_c}{\pi^2}\int_0^\infty dk\, k^2 \exp(-k^2/\beta_\infty^2)
\left\{ \frac{m}{\sqrt{m^2+k^2}} - \frac{m_Q}{\sqrt{m_Q^2+k^2}} \right\}, 
\end{eqnarray}
with $\beta_\infty\simeq 0.7$ GeV \cite{ms2004}. Notice that the quark condensate depends on the 
value of the current quark mass \cite{chpt,svz}.
For the physical value of the quark condensate, corresponding to the current quark mass $m=6$ MeV,   
we use $\langle \bar qq\rangle=-(240\pm 15\,\mbox{ MeV})^3$ \cite{jamin}. 
Eq.~(\ref{ms}) then gives $m_Q=220$ MeV, a typical value of the $u$ and $d$ constituent 
quark mass \cite{gi,melikhov}.   
In order to consider the chiral limit, $m\to 0$, the dependence of the quark 
condensate on the current quark mass should be taken into account. 
Setting $m=0$, and making use of the chiral quark condensate 
$\langle \bar qq\rangle_{m=0}\simeq -(230\pm 15\,\mbox{ MeV})^3$, 
Eq.~(\ref{ms}) gives the chiral constituent quark mass $m_Q^0=180$ MeV. 
This estimate has, however, rather an illustrative purpose: to find the true 
value of the chiral constituent quark mass in a given model,  
one should recalculate the meson spectrum and obtain $m_Q^0$ as the value for which the 
pion mass vanishes.  

\section{\label{5}Summary}
We demonstrated that the relativistic constituent quark picture based exclusively on  
constituent quarks is compatible with the chiral properties of QCD, 
if it has the following features: 
\begin{itemize}
\item[(i)] The axial coupling $g_A(s)$ of the constituent quark is a momentum-dependent quantity 
and is related to the pion $\bar QQ$ wave function by Eq.~(\ref{ga}).
\item[(ii)] The $\bar QQ$ potential saturates at large separations, $V(r\to\infty)\to \mbox{const}$. 
\end{itemize}
Under these conditions, a summation of the infinite number of diagrams describing 
constituent-quark soft interactions leads to the full axial current of the constituent 
quarks, conserved up to terms of order $O(M_\pi^2)$.

We defined the chiral point of the constituent quark model as those values of the parameters 
of the model (masses of the constituent quarks and couplings in the quark potential) for which 
the mass of the lowest pseudoscalar $\bar QQ$ bound state, $M_\pi$, vanishes. 
The chiral point of the constituent quark model corresponds to the spontaneously broken chiral 
limit of QCD: 
At the chiral point the full non-perturbative axial current of the constituent quarks is 
conserved (without explicit introduction of Goldstone degrees of freedom). 
The lowest part of the hadron spectrum has no other traces of chiral symmetry 
except for a massless pseudoscalar. Two important signatures of the spontaneously-broken 
chiral symmetry can be seen --- the decay constant of the massless pion is finite, whereas 
the decay constants of the excited massive pseudoscalars vanish.
%

We emphasize that the non-perturbative emergence of chiral symmetry in a model with 
only constituent quark degrees of freedom, reported in this paper, is qualitatively different from chiral 
symmetry in models which explicitly contain Goldstones along with constituent 
quark degrees of freedom: namely, the latter may be made chirally invariant for any value of the 
constituent quark mass, whereas in our approach the model is chirally symmetric only for a definite 
(non-vanishing) value of the constituent quark mass which leads to the massless ground-state pseudoscalar. 

Let us notice that the usual approximation, $g_A$ = const, may work reasonably for the calculation of 
most of the hadron properties beyond the chiral limit. However, within this approximation one gets wrong
properties of the excited pseudoscalars in the chiral limit. 

\vspace{0.8cm}

\noindent
{\it Acknowledgments.} 
We have the pleasure to express our gratitude to Berthold Stech for discussions which 
influenced to a great extent the concept of the paper. 
We would like to thank Vladimir Anisovich and Franz Schoberl 
for useful comments. D.~M.~was supported by the Austrian 
Science Fund (FWF) under project P17692.



\appendix

\section{\label{app}Analytic properties of the wave function in a potential model} 

To study the connection between the properties of the potential and the 
analytic structure of the wave function in momentum space, we start 
with the Schr\"odinger equation describing the interaction of two particles with mass 
$m$ in a relative $S$-wave by the potential $V(r)=\sigma r^a$:
\begin{eqnarray}
\left(-\frac{1}{m}\frac{d^2}{dr^2}+V(r)+\epsilon\right)r\Psi(r)=0.
\end{eqnarray}  
Here $\epsilon$ is the binding energy of the eigenstate with mass $M=2m-\epsilon$. 
For $a>0$, the wave function has the following behavior at large $r$: 
\begin{eqnarray}
\label{psir}
\Psi(r)\sim\frac{1}{r}\exp( -\sqrt{\sigma m} r^{1+a/2}). 
\end{eqnarray} 
The momentum-space wave function is obtained by Fourier transform and has the form 
($k=|\vec k|$): 
\begin{eqnarray}
\Psi(k)\sim  \int dr\,r\,\frac{\sin(kr)}{k} \Psi(r). 
\end{eqnarray}
To perform the analytic continuation in $\vec k^2$ to the unphysical negative values, 
we set $k=i\sqrt{z}$ and obtain
\begin{eqnarray}
\label{psiz}
\Psi(z)\sim  \int  dr\,r\,\frac{\exp(r\sqrt{z})-\exp(-r\sqrt{z})}{\sqrt{z}}\Psi(r). 
\end{eqnarray}
A singularity on the real axis of the variable $z$ at $z>0$ may emerge if the integral 
(\ref{psiz}) diverges at $r=\infty$ for some positive value of $z$. 
Evidently, for the wave function (\ref{psir}) with $a>0$ this does not happen: 
the integral (\ref{psiz}) is a regular function for all $z>0$. 
Therefore we conclude that for a potential rising for $r\to\infty$, the wave function in
the momentum space is a regular function on the real axis. 
Respectively, the vertex function $G(s)=(s-M^2)\Psi(s)$, with $s= 4m^2+4 k^2$, vanishes at $s=M^2$. 

The situation is different for a potential which saturates at large separations, 
$V(r\to\infty)=V_\infty$.
Then, the wave function behaves at large $r$ as 
\begin{eqnarray}
\label{psir2}
\Psi(r)\sim\frac{1}{r}\exp( -\mu r), \qquad \mu=\sqrt{m(V_\infty+\epsilon)}.
\end{eqnarray}
Setting $V_\infty=0$ and performing the Fourier transform, the momentum-space wave function takes the form  
\begin{eqnarray}
\label{psik2}
\Psi(\vec k^2)\sim  \frac{1}{\vec k^2+\mu^2} \sim \frac{1}{(s-M^2)}. 
\end{eqnarray}
Here we used the relation $s-M^2=4m^2+4k^2-(2m-\epsilon)^2=4(k^2+m\epsilon)+\epsilon^2
\simeq 4(k^2+m\epsilon)$ relevant for the non-relativistic treatment. 
The wave function (\ref{psik2}) has a pole at $s=M^2$, and thus the vertex function $G(s)=(s-M^2)\Psi(s)$ 
is finite at $s=M^2$. 
Notice that the location of the pole may be directly obtained from the Schr\"odinger
equation in the momentum-space representation, which for $r\to\infty$ reads
\begin{eqnarray}
\label{alg}
\frac{\vec k^2}{m} \Psi+V \Psi= -\epsilon  \Psi.  
\end{eqnarray}
For $V(r=\infty)=0$, we obtain the location of the pole at ${\vec k^2}=-m\epsilon$ solving 
Eq.~(\ref{alg}) as an algebraic equation. 

The generalization of the Schr\"odinger equation used for the calculation of the spectrum 
in relativistic quark models has a similar structure \cite{anis1}, 
\begin{eqnarray}
(\sqrt{s}+\hat V)\Psi = M \Psi, 
\end{eqnarray}
where $\hat V$ is the relativistic potential operator.  
If the potential vanishes at large separations, the wave
function $\Psi(s)$ has a pole at $s < 4m^2$ (below the two-particle threshold). 
The location of the pole can be found by solving the above equation as an algebraic equation. 
Then we find $\Psi(s)\sim 1/(s-M^2)$, and thus $G(s)=(s-M^2) \Psi(s)$ is finite for $s=M^2$.

\section{\label{aapp}Vector and scalar couplings of the constituent quarks}
We consider here vector and scalar couplings of the constituent quarks of different flavours 
defined as follows:  
\begin{eqnarray}
\langle Q_1Q_2|\bar q_1 \gamma_\mu q_2 |0\rangle_{\rm bare}&=&
g_V\bar Q_1 \gamma_\mu Q_2,
\nonumber\\
\langle Q_1Q_2 | \bar q_1 q_2|0\rangle_{\rm bare}&=&g_S \bar Q_1 Q_2, 
\end{eqnarray}
where $g_V(0)=1$ for the elastic vector current, and $g_V(0)$ is close to one for the weak current. 
In the first expression, we omit the possible structure $\bar Q_1\sigma_{\mu\nu}p^\nu Q_2$ 
\cite{ss} which is of no importance for our analysis. What can be said about the scalar coupling? 

Let us calculate the scalar and vector couplings of a scalar meson defined according to 
\begin{eqnarray}
\langle 0|\bar q_2 \gamma_\mu q_1|M(p)\rangle=p_\mu f_v,\nonumber\\  
\langle 0|\bar q_2 q_1|M(p)\rangle=f_s. 
\end{eqnarray}
By virtue of the QCD equations of motion, we have 
\begin{eqnarray}
\label{constraint2}
{(\bar m_1-\bar m_2)}f_s=M^2 f_v, 
\end{eqnarray}
where $M$ is the mass of the scalar meson, and $\bar m_1$, $\bar m_2$ are current quark masses. The  
corresponding constituent quark masses are denoted as $m_1$, $m_2$. 

The dispersion approach based on the constituent quark picture \cite{melikhov} 
leads to the following representations for the amplitudes:  
\begin{eqnarray}
\langle 0|\bar q_2 \gamma_\mu q_1|M(p)\rangle&=&{2p_\mu}
(m_2-m_1)\int ds g_V(s)\psi_s(s)\left(s-(m_1+m_2)^2\right) \frac{\rho(s)}{s},
\nonumber
\\
\langle 0|\bar q_2 q_1|M(p)\rangle&=&- \int ds g_S(s)\psi_s(s)\left(s-(m_1+m_2)^2\right) \rho(s). 
\end{eqnarray}
Making use of Eq.~(\ref{constraint2}) we find, for $s\ge 4m_Q^2$,  
\begin{eqnarray}
{(\bar m_1-\bar m_2)}\,g_S(s) ={(m_1-m_2)}\,g_V(s)\,\frac{M^2}{s}. 
\end{eqnarray}
Since constituent and current quark masses approximately obey the relation 
\begin{eqnarray}
{\bar m_1-\bar m_2}\simeq {m_1-m_2}, 
\end{eqnarray}
we find 
\begin{eqnarray}
g_S(s) \simeq g_V(s)\,\frac{M^2}{s}. 
\end{eqnarray}
No other constraints on the functional dependence of $g_V(s)$ and $g_S(s)$ emerge in this case.


\begin{thebibliography}{99}
\bibitem{chpt}J.~Gasser, H.~Leutwyler, Ann.~Phys.~(NY) {\bf 158}, 142 (1984). 
\bibitem{gi}S.~Godfrey, N.~Isgur, Phys.~Rev.~D {\bf 32}, 189 (1985). 
\bibitem{gromes}W.~Lucha, F.~F.~Schoberl, D.~Gromes, Phys.~Rep.~{\bf 200}, 127 (1991).
\bibitem{anisbook}V.~V.~Anisovich {\em et al.}, {\it Quark model and high-energy collisions} 
(World Scientific, Singapore 2004). 
\bibitem{amn} V.V.~Anisovich, D.~Melikhov, V.~A.~Nikonov, 
Phys.~Rev.~D {\bf 52}, 5295 (1995); 
Phys.~Rev.~D {\bf 55}, 2918 (1997). 
\bibitem{ms}M.~Beyer, D.~Melikhov, Phys.~Lett.~B {\bf 436}, 344 (1998); \\
D.~Melikhov, B.~Stech, Phys.~Rev.~D {\bf 62}, 014006 (2000).
\bibitem{silvano}R.~Petronzio, S.~Simula, G.~Ricco, Phys.~Rev.~D {\bf 67}, 094004 (2003); \\
S.~Simula, Phys.~Lett.~B {\bf 574}, 189 (2003).

\bibitem{mg}A.~Manohar, H.~Georgi, Nucl.~Phys.~B {\bf234}, 189 (1984);
\bibitem{se}U.~Ellwanger, B.~Stech, Phys.~Lett.~B {\bf 241}, 409 (1990); Z.~Phys.~C {\bf 49}, 683 (1991). 

\bibitem{lucha}W.~Lucha, D.~Melikhov, Phys.~Rev.~D {\bf 73}, 054009 (2006). 
\bibitem{svz}M.~A.~Shifman, A.~I.~Vainshtein, V.~I.~Zakharov, Nucl.~Phys.~B {\bf 147}, 385 (1979).  
\bibitem{krassnigg}A.~Holl, A.~Krassnigg, C.~D.~Roberts, Phys.~Rev.~C {\bf 70}, 042203(R) (2004). 
\bibitem{gmor}M.~Gell-Mann, R.~J.~Oakes, B.~Renner, Phys.~Rev. {\bf 175}, 2195 (1968). 
\bibitem{anisovich}V.~V.~Anisovich {\em et al.}, Nucl.~Phys.~A {\bf 544}, 747 (1992);\\ 
V.~V.~Anisovich {\em et al.}, J.~Phys.~G {\bf 28}, 15 (2002). 
\bibitem{melikhov}D.~Melikhov, Phys.~Rev.~D {\bf 53}, 2460 (1996); Phys.~Rev.~D {\bf 56}, 7089 (1997); 
EPJdirect {\bf C4}, 2 (2002) [hep-ph/0110087].
\bibitem{berthold}
D.~Melikhov, B.~Stech, hep-ph/0606203. 
\bibitem{ss}F.~Cardarelli {\em et al.}, Phys.~Lett.~B {\bf 332}, 1 (1994); 
Phys.~Lett.~B {\bf 359}, 1 (1995); Phys.~Rev.~D {\bf 53}, 6682 (1996); \\
F.~Cardarelli {\em et al.}, Phys.~Lett.~B {\bf 357}, 267 (1995); 
Phys.~Lett.~B {\bf 371}, 7 (1996); Phys.~Lett.~B {\bf 397}, 13 (1997); \\ 
F.~Cardarelli, B.~Pasquini, S.~Simula, Phys.~Lett.~B {\bf 418}, 237 (1998); \\ 
F.~Cardarelli, S.~Simula, Phys.~Rev.~C {\bf 62}, 065201 (2000); \\
S.~Simula, in Proc. of the NSTAR 2001 Workshop on {\em The Physics of Excited 
Nucleons}, Mainz, Germany, 7-10 Mar 2001, D.~Drechsel and L.~Tiator, eds., 
World Scientific Publishing (Singapore, 2001), pg. 135 [arXiv: nucl-th/0105024].

\bibitem{weinberg}S.~Weinberg, Phys.~Rev.~Lett. {\bf 65}, 1181 (1990); 
Phys.~Rev.~Lett. {\bf 67}, 3473 (1991). 
\bibitem{charm}D.~Melikhov, M.~Beyer, Phys.~Lett.~B {\bf 452}, 121 (1999); \\ 
D.~Melikhov, O.~Pene, Phys.~Lett.~B {\bf 446}, 336 (1999).
\bibitem{ms2004}D.~Melikhov, S.~Simula, Eur.~Phys.~J.~C {\bf 37}, 437 (2004);
\bibitem{jamin}M.~Jamin, Phys.~Lett.~B {\bf 538}, 71 (2002). 
\bibitem{anis1}V.~V.~Anisovich {\em et al.}, Phys.~At.~Nucl.~{\bf 68}, 1573 (2005). 
\end{thebibliography}
\end{document}